\documentclass[twocolumn,showpacs,aps,prb,floatfix]{revtex4}
\usepackage{amssymb}

\usepackage{graphicx}

\begin{document}

\title{Anomalous magnetotransport in (Y$_{1-x}$Gd$_{x}$)Co$_{2}$
alloys: interplay of disorder and itinerant metamagnetism.}
\author{A. T. Burkov, A. Yu. Zyuzin}
\affiliation{A. F. Ioffe Physico-Technical Institute, Russian
Academy of Sciences, Saint-Petersburg 194021, Russia.}
\author{T. Nakama, K. Yagasaki}
\affiliation{Physics Department, University of the Ryukyus, Okinawa 903-0213, Japan}
\date{\today}

\begin{abstract}
New mechanism of magnetoresistivity in itinerant metamagnets with
a structural disorder is introduced basing on analysis of
experimental results on magnetoresistivity, susceptibility, and
magnetization of structurally disordered alloys
(Y$_{1-x}$Gd$_{x}$)Co$_{2}$. In this series, YCo$_{2}$ is an
enhanced Pauli paramagnet, whereas GdCo$_{2}$ is a ferrimagnet
(T$_{\rm c}$=400 K) with Gd sublattice coupled
antiferromagnetically to the itinerant Co-3d electrons. The alloys
are paramagnetic for $x < 0.12$. Large positive magnetoresistivity
has been observed in the alloys with magnetic ground state at
temperatures T$<$T$_{\rm c}$. We show that this unusual feature is
linked to a combination of structural disorder and metamagnetic
instability of itinerant Co-3d electrons. This new mechanism of
the magnetoresistivity is common for a broad class of materials
featuring a static magnetic disorder and itinerant metamagnetism.
\end{abstract}

 \maketitle

\section{Introduction}
Interplay of structural disorder and magnetic interactions opens a
rich field of new physical phenomena. Among them are the actively
discussed possibility of disorder-induced Non-Fermi Liquid (NFL)
behavior near a magnetic Quantum Critical Point (QCP) as well as a
broader scope of effects of disorder on magnetotransport.
\cite{Hertz76,Millis93,Varma2001} Structurally disordered alloys
Y$_{1-x}$Gd$_{x}$Co$_{2}$ are quasi-binary solid solutions of
Laves phase compounds YCo$_2$ and GdCo$_2$. The compounds belong
to a large family of isostructural composites RCo$_2$. YCo$_2$ is
an enhanced Pauli paramagnet whose itinerant Co-3d electron system
is close to magnetic instability. In external magnetic field of
about 70~T this system undergoes a metamagnetic transition into
ferromagnetic (FM) ground state.\cite{Goto89} GdCo$_2$ is, on the
other hand, a ferrimagnet with a Curie temperature of 400~K in
which the spontaneous magnetization of 4f moments is anti-parallel
to the induced magnetization of the Co-3d band. Compounds of this
family and their alloys provide a convenient ground for
experimental studies of magnetotransport phenomena. The electronic
structure in the important for the transport vicinity of the Fermi
energy is composed mainly of Co-3d states and is, to the first
approximation, the same for all compounds of RCo$_2 $ family. It
has been found that the main contribution to the resistivity of
RCo$_2$ comes from the scattering of conduction electrons on
magnetic fluctuations due to strong s--d exchange coupling,
\cite{Gratz95} therefore the transport properties are expected to
be especially sensitive to the magnetic state of the sample.
GdCo$_2$ occupies a special place in RCo$_2$ family since Gd 4f
magnetic moment has no orbital contribution and, therefore
crystal-field effects are not important for this compound.

The experimental results on the transport properties of
Y$_{1-x}$Gd$_{x}$Co$_{2}$ alloys has been published partly in our
previous article. \cite{Nakama2001} Here we analyze these and new
experimental results in order to reveal the physical mechanism of
anomalous megnetotransport properties observed in the alloys. In
this paper we will discuss the magnetotransport properties of the
FM alloys. The properties of paramagnetic alloys will be published
elsewhere.

\section{Experimental}
Samples of Y$_{1-x}$Gd$_{x}$Co$_{2}$ were prepared from pure
components by melting in an arc furnace under a protective Ar
atmosphere and were subsequently annealed in vacuum at 1100~K for
about one week. An X-ray analysis showed no traces of impurity
phases. A four--probe dc method was used for electrical
resistivity measurements. Magnetoresistivity (MR) was measured
with longitudinal orientation of electrical current with respect
to the magnetic field. The size of the samples was typically about
1$\times $ 1$ \times $10~mm$^{3}$. Magnetization was measured by a
SQUID magnetometer for samples from the same ingot as that used
for the resistivity and AC susceptibility measurements.
\section{Experimental results}
\begin{figure}[h]
\includegraphics[width=1.0\linewidth]{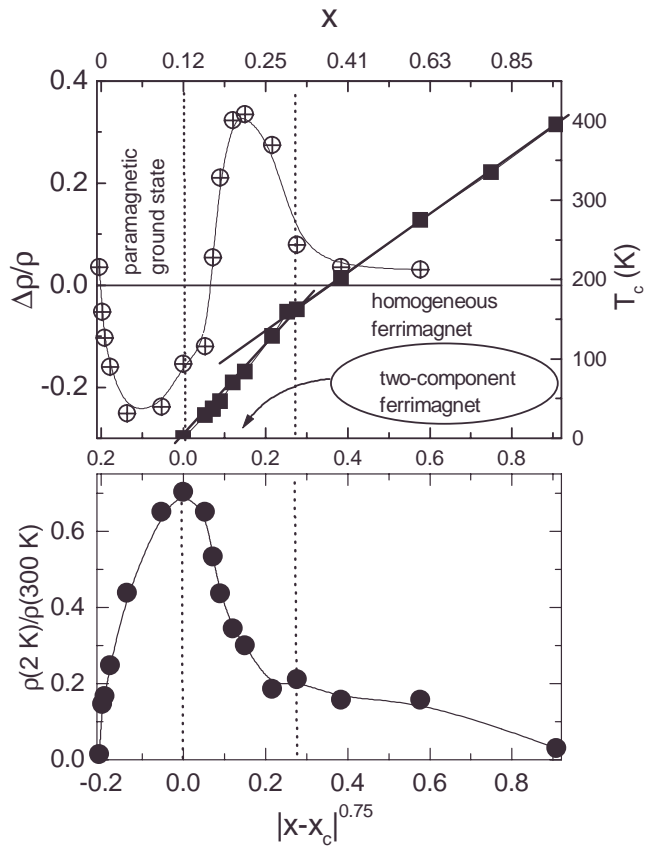}
\caption{The upper panel shows the ordering temperature T$_{\rm
c}$ $\blacksquare $ (right y-axis), and the MR $\bigoplus $ (left
axis) of the Y$_{1-x}$Gd$_{x}$Co$_{2}$ system \cite{Nakama2001}.
The MR was measured at T = 2~K in magnetic field of 15~T. The
dotted vertical lines indicate phase boundaries at zero
temperature. The lower panel displays normalized resistivity
$\frac{\rho(2~K)}{\rho(300~K)}$.} \label{PasDiag}
\end{figure}

\begin{figure}[hbt]
\includegraphics[width=1.0\linewidth]{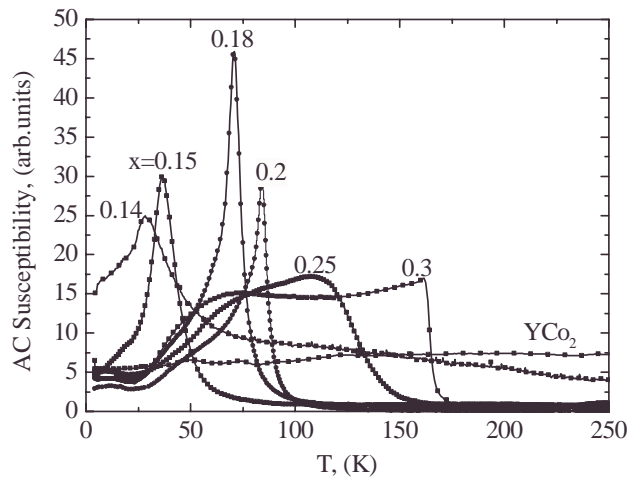}
\caption{The ac susceptibility of the Y$_{1-x}$Gd$_{x}$Co$_{2}$ alloys.
Note, the experimental data for YCo$_2$ and for the alloy with $x=0.14$ are
multiplied by factor 20.}
\label{Sus}
\end{figure}

The magnetic phase diagram of the Y$_{1-x}$Gd$_{x}$Co$_{2}$ system
inferred from the transport and magnetic measurements
\cite{Nakama2001} is shown in Fig.~\ref{PasDiag}.
Curie temperature $T_{\mathrm{c}}$ decreases with increasing
content of Y and eventually drops to zero. A precise determination
of the critical concentration $x_c$ which separates the
magnetically ordered ground state and the paramagnetic region is
difficult, since on the onset of the long range order its
signatures in the magnetic and transport properties are very weak.
The first firm evidence of the long range order are found for
alloy with x=0.14 in ac susceptibility at T=27 K, Fig.~\ref{Sus}.

Quantum critical scaling theory predicts that when Curie temperature of a FM system
continuously depends on an external parameter $x$, this dependence
is expressed as \cite{Millis93}: $$ T_{\rm c}\varpropto \left|
x-x_{\rm c}\right| ^{\frac{z}{d+z-2}}$$ with critical index $z=3$
for a FM system of spatial dimension $d=3.$ The experimental
T$_{\rm c}$ vs. $x$ dependency does follow this relation, but with
additional kink at $x=x_{\rm t}$. A possible origin of this kink
will be discussed later. Linear extrapolation of the phase
separation line on the phase diagram Fig. \ref{PasDiag} to $T_{\rm
c} =0$ gives as the critical concentration $x_c=0.12$. We do not
claim however that QCP exists in this alloy system. Direct
experimental verification that T$_{\rm c}$ $ \rightarrow$ 0 as $x$
approaches $x_{\rm c}$ from magnetically ordered state is
difficult for a disordered alloy system.

\begin{figure}[h]
\includegraphics[width=1.1\linewidth]{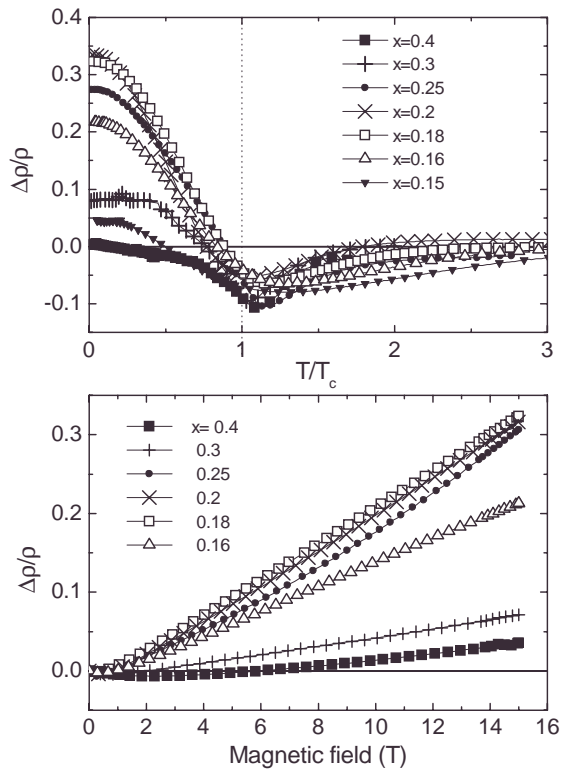}
\caption{The upper panel shows the temperature dependence of the
MR of the Y$_{1-x}$Gd$_{x}$Co$_{2} $ alloys, measured in field of
15~T. Large positive MR of FM alloys ($x\leq0.3$) is observed at
low temperatures. The field dependencies of MR, measured at T=2~K,
are presented on the lower panel.} \label{DRvsT}
\end{figure}

The very surprising result is the positive MR in the FM phase at
low temperatures, see Figs.~\ref{PasDiag} and \ref{DRvsT}. The
well known theoretical result for MR of a localized moment
ferromagnet was derived long ago by Kasuya and De Gennes.
\cite{Kasuya56} As it follows from their theory, MR of a metallic
ferromagnet should be negative, having a maximum absolute value at
Curie temperature, and approaching zero as T $\rightarrow$ 0, and
in the limit of high temperatures. Qualitatively this behavior has
been supported by experiment, as well as by later more detailed
theoretical calculations. The present experimental results are in
a qualitative agreement with this theoretical behavior only for
alloys with large Gd content ($x \geqslant 0.4$)
(Fig.~\ref{DRvsT}). MR of the FM alloys with composition
$0.3>x>0.14$ fundamentally differs from this theoretical behavior.
Let us note that this composition range falls into the region of
the phase diagram between the paramagnetic phase and the
additional phase boundary indicated by the kink in the T$_c$ vs
$x$ dependency, see Fig. \ref {PasDiag}. MR of these alloys is
positive below Curie temperature and is very large.

The known mechanisms of a positive MR can not explain the
experimental data. A rough estimate of Lorenz force-driven MR one
can get from a comparison with the MR of pure YCo$_2$.
\cite{Burkov98} In the most pure samples of YCo$_2$ (with residual
resistivity of about 2~$\mu \Omega $cm) the Lorenz force-driven
positive contribution to the total MR does not exceed 5\%. On the
other hand the resistivity of the FM alloys at low temperatures
falls into the region from 30 to 100~$\mu \Omega $ cm, i.e. at
least one order of magnitude larger than the resistivity of pure
YCo$_2$. Therefore, according to Koehler's rule, this mechanism
can give MR of only about 0.5\%, this has to be compared with the
experimental MR of almost 40\%.

Weak localization effect is known to give positive MR. However our
estimates show that this mechanism can give a contribution which
is at least two order of magnitude smaller than the observed MR.
\section{Discussion}
The key for understanding the mechanism of the positive MR is a
combination of strong dependence of the magnetic susceptibility
$\chi$ of metamagnetic Co-3d subsystem on the effective magnetic
field and of the structural disorder in the R - sublattice of the
alloys. In case of GdCo$_2$ as well as in the case of other
magnetic RCo$_2$ compounds with heavy R-elements, the 4f--3d
exchange interaction is described by introducing an effective
field which acts on 3d electrons as: $$B_{\rm eff}=n_{\rm
fd}M_{\rm f}-B$$ where B is external field, M$_f$ is the uniform
magnetization of R - sublattice, and $n_{\rm fd}$ - is the f--d
coupling constant (in case of GdCo$_2$ $n_{\rm fd}\approx 50$~T/f.u.$\mu_{\rm B}$\cite{Goto2001}).
In Y$_{1-x}$Gd$_{x}$Co$_{2}$ alloys the Gd
moments are randomly distributed over the R-sites of the crystal
lattice. Therefore, the effective field acting on 3d electrons
depends on the local distribution of Gd moments and is therefore a
random function of coordinate. This random field can be
characterized by a distribution function $P\{B_{\rm eff}(r)\}$.
The spatially fluctuating effective field induces an inhomogeneous
magnetization of Co-3d electron system: $$m(r)=\chi \left( B_{\rm
eff}\right) B_{\rm eff}\left( r\right).$$ Therefore even at zero
temperature in the ferromagnetic ground state, there are two kind
of static magnetic fluctuations in the system: i. $M_{\rm f}(r)$;
ii. $m(r)$. These fluctuations give an additional contribution to
the resistivity. Since at T=0~K in ferromagnetic phase the 4f
magnetic moments are saturated the corresponding contribution to
the resistivity does not depend on external magnetic field.  On
the other hand, the 3d magnetic moment, as we will see later, is
not saturated even in the ferromagnetic ground state. Therefore
the 3d magnetization does depend on the external field. However,
as long as 3d susceptibility is field independent and uniform, the
external magnetic field will change only the mean (non-fluctuating
part) value of the magnetization, whereas the magnitude of the
fluctuations of $m$ and corresponding contribution to the
resistivity remain unchanged. However actually, the 3d system is
close to the metamagnetic instability. Therefore the 3d
susceptibility $\chi $ is field dependent and gives rise to static
magnetic fluctuations which are dependent on magnetic field
resulting in non-zero magnetoresistivity.

For a qualitative analysis of this new mechanism of MR we make the
following assumptions.\\ 1.We consider the system only in its
ground state.\\ 2.Correlations between potential and
spin-dependent scattering are neglected.\\ 3.Only the
contributions which may be strongly dependent on external magnetic
field are retained, e.g. we do not include the effects related to
a change of d-density of states in magnetic field, and
contributions due to potential scattering and scattering on 4f
moments. In fact only the contribution related to the scattering
on fluctuations of Co-3d magnetization will be considered.\\

\begin{figure}[h]
\includegraphics[width=1.0\linewidth]{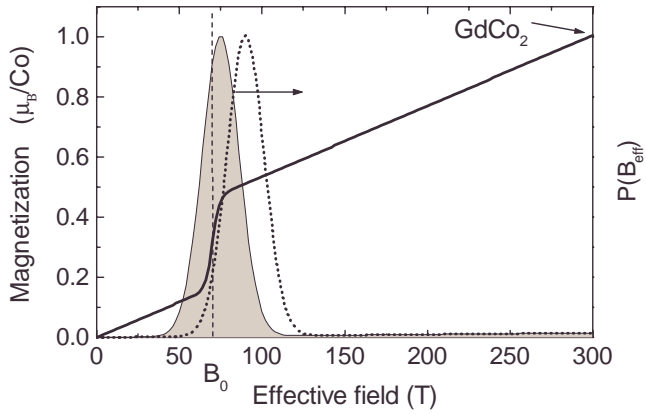}
\caption{Dependence of the magnetization of Co--subsystem in
RCo$_2$ compounds vs effective magnetic field (left axis)
\cite{Goto2001} (shown schematically). The dotted line is a
schematic representation of the distribution function $P(B_{\bf
eff})$ (right axis), the shaded area indicates the position of
$P(B)$ in external field of 15~T.} \label{Co-magnet}
\end{figure}

\begin{figure}[h]
\includegraphics[width=1.0\linewidth]{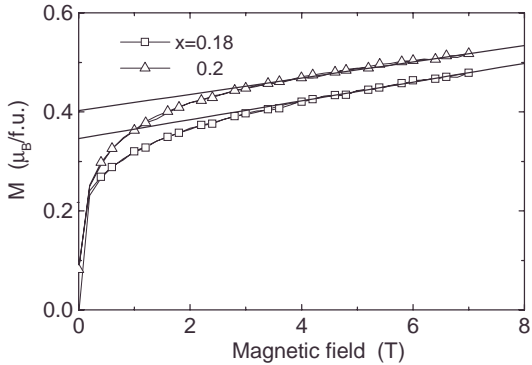}
\caption{The magnetization of the Y$_{1-x}$Gd$_{x}$Co$_{2}$ alloys
vs magnetic field.} \label{Magnz}
\end{figure}

Figure~\ref{Co-magnet} shows schematically the dependency of the
Co-3d magnetization of RCo$_{2}$ compounds on effective magnetic
field \cite{Goto2001} and the distribution function  $P\{B_{\rm
eff}(r)\}$. The metamagnetic transition is indicated by the rapid
increase of the magnetization around 70~T.
For the further discussion it is important to have an estimate of
the magnitude of the fluctuations of the effective field, i.e. the
width of the distribution function  $P\{B_{\rm eff}(r)\}$. We can
get a hint considering experimental data on field (0 to 7~T)
dependency of the magnetization, Fig.\ref{Magnz}.

Two important points follow from these data: i. there is a
non-saturating para-process in these field dependencies above
about 2 T; ii. the estimated susceptibility of this para-process
($\approx$ 0.016~$\mu_B$/T) is larger than the susceptibility of
3d system below and above metamagnetic transition ($\approx $
0.002~$\mu_B$/T), however it is smaller than the susceptibility in
the transition region ($\approx$ 0.04~$\mu_B$/T).\cite{Goto2001}
This indicates, that the scale of the fluctuations is larger than
the width of the metamagnetic transition. Therefore we can not
treat the fluctuating part of the effective field as a
perturbation and have to resort to a phenomenological analysis.

\begin{figure}[h]
\includegraphics[width=1.0\linewidth]{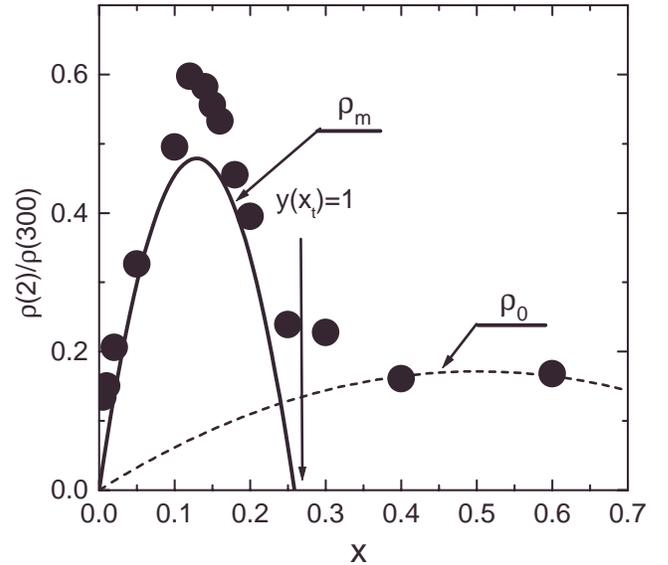}
\caption{Schematic of contributions to the low-temperature
resistivity of Y$_{1-x}$Gd$_{x}$Co$_{2}$ alloys. $\bullet $
-experimental normalized resistivity measured at T=2~K in field of
15~T. Solid line represents schematically $\rho _m$, broken line -
$\rho _0$.} \label{decom}
\end{figure}

The distribution function $P(B_{\bf eff})$ describing the
fluctuating effective field depends on the alloy composition. For
diluted alloys ($x \approx 0$) the most probable value $B _{\rm
av}$ of $B_{\rm eff}$ is close to zero. As $x$ increases $B _{\rm
av}$ shifts to higher values and in a certain range of the
concentrations, the function will have essentially non-zero weight
for both $B_{\bf eff} < B_0$ and $B_{\bf eff} > B_0$, see
Fig.~\ref{Co-magnet}. In this case there shall be regions with low
and with high 3d magnetization in the sample. The resistivity
resulting from this static disorder in 3d magnetization can be
expressed as: $$\rho_m = \rho_{sd}\cdot y\left(1-y\right).$$ Where
$ y=\int_{B_0}^{\infty} P\left(B_{\rm eff}\right)dB$ is the volume
fraction of the high magnetization component. Parameter $y$
depends on the alloy composition $x$, and on external magnetic
field $B$. In zero field, point $y=0$ corresponds to $x=0.$ As the
content of Gd increases $B _{\rm av}$ shifts to larger effective
fields, and finally, at some alloy composition $x_t$, it becomes
larger than $B_0$, i.e. at this composition almost whole volume is
occupied by the high magnetization component, i.e. $y\approx 1$.
With a further increase of Gd content the mean magnetization of Co
should increase with a smaller rate, determined by the slope of
the $m(B)$ dependency above $B_0$, however $y=1$ in this region.
According to this scenario $\rho_m$ will increase with $x$ at
first, reach a maximum value at $x$ which corresponds to $y=0.5$
and will decrease with further increase of $x$ approaching to zero
at $x\approx x_t$. We believe that $x_t$ corresponds to the kink
on the phase diagram, Fig.~\ref{PasDiag}.  The expected variation
of $\rho _m$ with the alloy composition is schematically shown in
Fig. \ref{decom}. The total experimental resistivity includes
additionally contributions coming from potential scattering and
from scattering on 4f moments ($\rho_0$), which both are
proportional to $x(1-x)$ and also are depicted in Fig.
\ref{decom}. For comparison the experimental resistivity is shown
in this picture too. We present here the resistivity measured at
T=2~K in field of 15~T to exclude the contributions related to
spin-flip scattering.

The experimental low-temperature resistivity shows the expected
variation with $x$ (one needs to keep in mind that the relation
between $x$ and $y$ is in general non-linear, especially around
magnetic phase boundary $x_c$). The resistivity attains the
maximum value of about 100~$\mu \Omega\, cm$ in the region, which
corresponds to maximum static magnetic disorder at $y \approx 0.5$
(room temperature resistivity of the alloys weakly depends on $x$
and is about 150~$\mu \Omega\, cm$). About the same value was
obtained for the high temperature limit (maximum magnetic
disorder) of magnetic part of the resistivity, arising from
scattering on 3d temperature-induced magnetic fluctuations in
YCo$_2$.\cite{Gratz95} This suggests that the main part of the
experimental low temperature resistivity at $x<x_t$ originates
from the scattering on the magnetic fluctuations of 3d
magnetization, i.e. is identical to $\rho _{\rm m}$.

Let us show that $\rho _{m}$ is of order of the spin-fluctuation
contribution to resistivity at large temperatures, i.e. is of
order of 100~$\mu \Omega$cm. In case of YCo$_{2}$ the
spin-fluctuation contribution is the most important and is nearly
independent of temperature above about 200~K.
Hamiltonian of s-d
exchange interaction is given by

$$H_{sd}=G\int
d\mathbf{rs}(\mathbf{r})\mathbf{S}_{d}(\mathbf{r})$$

here $\mathbf{s}(\mathbf{r})$ and $\mathbf{S}_{d}(\mathbf{r})$ are
spin density of s- and d- electrons, correspondingly.\footnote{Hereafter we take
$\mu _B \equiv 1$.}

Spin fluctuation contribution to resistivity has form
\cite{Ueda75}:

\begin{equation}
\rho =\frac{3m}{4ne^{2}}G^{2}N_{s}\frac{1}{T}\int\limits_{0}^{1}dqq^{3}\int%
\limits_{-\infty }^{\infty }\frac{d\omega \omega }{\sinh
^{2}\left( \omega /2T\right) }Im\chi \left(
q\frac{2k_{F}}{k_{F}^{\ast }},\omega \right) \label{Ueda}
\end{equation}

where\ $N_{s}$ is density of states of s-electrons, $\frac{k_{F}}{%
k_{F}^{\ast }}$ is ratio of Fermi momentum of s- and d- electrons.
Dynamic susceptibility $\chi \left( q,\omega \right)$ is given by
the equation:

$$\chi ^{-1}\left( q,\omega \right) =\chi ^{-1}\left( q\right)
\left( 1-i\omega /\Gamma _{q}\right) $$

here $\Gamma _q$ is damping of the spin-fluctuations, whereas static nonlocal susceptibility
$\chi ^{-1}\left( q\right)$ is given by

$$\chi ^{-1}\left( q\right) =\chi ^{-1}+Aq^{2}/N_{d}$$

with  $\chi =N_{d}/\zeta(T)$ , where  $\zeta(T)$ is
inverse Stoner factor, renormalized by spin fluctuations, $N_{d}$
is density of states of d-electrons, and $A < 1$ is a dimensionless constant.

At large temperatures
$T>\Gamma _q$ the expression (\ref{Ueda})
reduces to:

\begin{equation}
\rho =\frac{3\pi
m}{ne^{2}}G^{2}N_{s}T\int\limits_{0}^{1}dqq^{3}\chi \left( q
\frac{2k_{F}}{k_{F}^{\ast }}\right) \simeq \frac{3\pi m}{4ne^{2}}
G^{2}N_{s}T\chi \label{rred}\end{equation}

The last equality is valid when $2Ak_{F}/k_{F}^{\ast }<\zeta(T)<1$
which must be the case for YCo$_{2}$.  Note that neglecting
momentum dependence of susceptibility means that s-electrons see
the d-spin fluctuations as point scatterers.

Scattering of conducting s -- electrons by the static random
distribution of spin density of d-electrons we can estimate in the
following way.

Due to random distribution of magnetic moments of Gd, s-electrons
experience scattering by $$\frac{G}{2}\left\langle
S_d\left(\mathbf{r}\right)\right\rangle= \frac{G}{2}\int
\chi_{m}(\mathbf{r-r}^{\prime
})\delta B_{eff}(\mathbf{r}^{\prime })$$ random potential. Here $\chi _{m}(%
\mathbf{r-r}^{\prime })$ is nonlocal susceptibility near
metamagnetic transition. We estimate correlation function of
fluctuating effective field as $$\left\langle \delta
B_{eff}(\mathbf{r})\delta B_{eff}(\mathbf{r}^{\prime
})\right\rangle =\left(2 S_{\rm Gd}n_{\rm fd}\right)^2\delta \left(\mathbf{r-r}^{\prime }\right)x\left(
1-x\right) a^{3}.$$ Here $a^{3}$ is volume of the formula unit. In
Born approximation the corresponding contribution to the
resistivity is given by the expression:

\begin{equation}
\rho _{m}=\frac{m}{ne^{2}}G^{2}N_{s}\left(2 S_{\rm Gd}n_{\rm fd}\right)^2x\left( 1-x\right)
a^{3}\int\limits_{0}^{1}dqq^{3}\chi _{m}^{2}\left( q\frac{2k_{F}}{%
k_{F}^{\ast }}\right)
\label{stat}
\end{equation}

Both expressions (\ref{Ueda}) and (\ref{stat}) give the
contributions to the resistivity due to scattering on spin
fluctuations, however in the first case they are of thermal
origin, whereas in the second case the fluctuations are due to
randomness of the effective field.

Assuming that nonlocality of the susceptibility is not important
we obtain from (\ref{rred}) and (\ref{stat}):

$$\frac{\rho _{m}}{\rho }=\frac{\left(2 S_{\rm Gd}n_{\rm fd}\right)^2x\left( 1-x\right) a^{3}\chi _{m}^{2}%
}{3\pi T\chi }$$

Using experimental results for $\chi _{\rm m}$ and $\chi$
\cite{Goto89,Goto2001,Burzo72} we find $\frac{\rho _{m}}{\rho }$
in the range  from 0.5 to 3, i.e. the resistivity caused by the
static magnetic fluctuations is of the same order as the
temperature--induced spin fluctuation resistivity. The uncertainty
is mainly due to determination of $\chi _{\rm m}$ near the
metamagnetic transition. Taking $\chi _{\rm m}$ as the
susceptibility of YCo$_2$ at B=70~T (at the field of metamagnetic
transition)\cite{Goto89} gives the upper bound for $\frac{\rho
_{m}}{\rho }$. Whereas $\chi _{\rm m}$ for the disordered alloy of
$x$=0.18, estimated from our results on M(B), Fig. \ref{Magnz}
gives the lower bound.

In external magnetic field, the effective field $B_{\bf eff}$
decreases \footnote{The effective field decreases for
antiferromagnetic 4f-3d exchange, for ferromagnetic exchange (like
that in RCo$_2$ compounds with light R-elements) the effective
field increases with the external field.}, therefore $y$ also
decreases. Depending on the value of $y_0$ -- the volume fraction
in zero field, $\rho _{\rm m}$ will either increase or decrease,
resulting in positive or negative magnetoresistivity: for
$0.5<y_0<1$ it will be positive, whereas for $0<y_0<0.5$ we will
have a negative MR. In agreement with the model, the experimental
MR is positive at $0.15<x<x_t$ and quickly decreases at $x > x_t =
0.3$ where $y \approx 1$. Nearly linear field dependencies of MR,
observed for $ x<0.3$, see Fig.~\ref{DRvsT}, implies that the
width of $P(B_{\rm eff})$ in this composition range is larger than
our experimental field limit of 15~T. The region $y < 0.5$ almost
coincides with the paramagnetic region of the phase diagram. The
model predicts negative MR for this region, and this prediction
agrees with the experimental result. The model also gives a
satisfactory description of the residual resistivity behaviour in
this region, see Fig.~\ref{decom}. This agreement suggests that at
$x < x_c$ the system is actually in spin-glass state. In this
region additional essential contributions to MR are present.
First, in the paramagnetic region there is a negative MR due to
suppression of magnetic disorder in 4f magnetic moment system by
external magnetic field. This negative contribution may be the
reason why the cross-over point from the positive to negative
magnetoresistivity does not coincide with the maximum of
resistivity, see Fig.~\ref{PasDiag}. Secondly, there can be
additional contributions, both positive and negative, near to
zero-temperature magnetic phase boundary due to closeness to QCP.

\begin{figure}[h]
\includegraphics[width=1.0\linewidth]{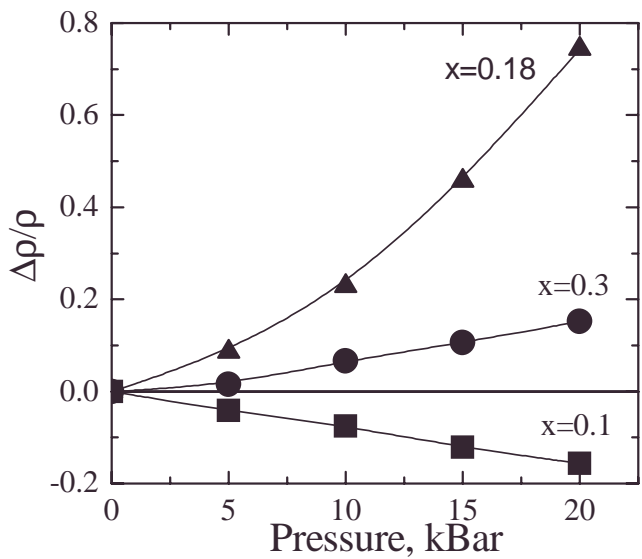}
\caption{The resistivity of the Y$_{1-x}$Gd$_{x}$Co$_{2}$ alloys
vs external hydrostatic pressure at T = 2~K.} \label{Press}
\end{figure}

An independent test of the model is based on the observation that
the critical magnetic field $B_0$ of the metamagnetic transition
in 3d subsystem increases under external hydrostatic pressure (P).
\cite{Saito99,Yamada95} Therefore, basing on the model, we expect
that the resistivity of the alloys with the composition left of
the resistivity maximum (see Fig.~\ref{PasDiag}) will decrease
with pressure, whereas for the alloys right of the maximum it will
increase with the increasing pressure. The experimental results
for three alloy compositions are shown in Fig.~\ref{Press}.
The sign of the pressure effect is in agreement with the model
prediction: the resistivity decreases with pressure for $x$=0.1,
whereas it increases with P for $x$=0.18 and $x$=0.3. Moreover,
there is a good scaling of pressure and magnetic field
dependencies of the resistivity,  Figs.~\ref{DRvsT} and
\ref{Press} (P$\rightarrow\alpha $B) with scaling parameter
$\alpha$ which is close to literature data on the pressure
dependence of B$_0$: $\frac{dB_0}{dP}\approx
1.5$~T/kBar.\cite{Saito99,Yamada95}

\section{Conclusion}
It has been found that at low temperatures there is a large
contribution to the resistivity related to scattering on magnetic
fluctuations in metamagnetic itinerant 3d system, induced by
fluctuating effective field of 4f moments. Large positive MR,
found in the FM Y$_{1-x}$Gd$_x$Co$_2$ alloys and strong pressure
dependence of the resistivity are explained as arising from a
combination of static magnetic disorder and strong magnetic field
dependence of magnetic susceptibility.

We want to emphasize that this mechanism of resistivity (and of
MR) is not material specific, rather it should be common for a
broad class of disordered itinerant metamagnets with strong
coupling of conduction electrons to the magnetic fluctuations. In
Y$_{1-x}$Gd$_x$Co$_2$ the relevant disorder originates from random
distribution of d--f exchange fields, however a similar effect
should arise when there is a random distribution of local
susceptibilities (a corresponding treatment for Kondo systems was
recently developed in \cite{Wilhelm2001}).
Positive MR observed in FM alloys Y(Co$_{1-x}$Al$_x$)$_2$ \cite%
{Nakama2000} and in Er$_{x}$Y$_{1-x}$Co$_2$ \cite{Hauser2001} may
be explained by this mechanism.
\section*{ACKNOWLEDGMENT}
This work is supported by grants 02-02-17671 and 01-03-17794 of
Russian Science Foundation, Russia. We thank Dr. P. Konstantinov
for stimulating discussions.

\end{document}